# Quantum Cattaneo wave equation for ultra-short laser pulses interaction with electron and nucleon gases

J.Marciak-Kozłowska[1], M.Kozłowski[2,*]

[1] Institute of Electron Technology, Warsaw, Poland,
[2,*] Physics Department, Warsaw University, Warsaw, Poland

*Corresponding author, e-mail: miroslawkozlowski@aster.pl

Abstract

In this paper the quantum Cattaneo wave equation for ultra-short laser pulses interaction with matter is obtained. The explicit formulae for electron and nucleon gases are presented

Introduction

From the theoretical point of view it seems quite obvious that diffusive processes can not take place with infinite velocity inside matter, since this fact would violate causality in the special relativity framework. However, the standard diffusion equation is based on Fick's law [1] in the case of mass transport, and on Fourier's law [2] in the case of heat conduction. Some other quite common constitutive laws of mathematical physics establish a similar time-independent functional relation between the flux and the spacial gradient of the state variable. This is the case of Ohm's law in electricity [3], of Darcy's law in fluid motion within porous media [4], etc. When combined with the adequate conservation principle (i.e.: the continuity equation), these constitutive laws lead to a pure parabolic mathematical model that predicts an infinite speed of propagation for the mass or energy being transported. For these reasons, the infinite speed paradox underlies many mathematical models that are frequently used in computational engineering and science. It happened to be ironical that the first comprehensive derivation for the standard diffusion equation was given by Einstein himself. This issue is just ignored in a number of applications in which the standard linear parabolic model is supposed to be accurate enough for practical purposes, although the simple idea of mass or energy being transported at infinite speed is disturbing. However, in some other applications it could be mandatory to take into account the wave nature of diffusive processes to perform accurate predictions].For the above stated reasons, a considerable effort has been historically devoted to remove the infinite speed paradox from the standard diffusion equation. Basically, two kind of approaches have been presented: on one side, some non-linear constitutive laws have been proposed, thuskeeping the parabolic nature of the model [; on the other, some time-dependent constitutive laws have been proposed, what leads to a new class of hyperbolic-type wave models, as the one first proposedby Cattaneo [6].

Cattaneo equation is the classical wave equation for the temperature field *T(x,t)*. In this paper we develope the generalized quantum Cattaneo equation for heat transport phenomena induced by attosecond ( $10^{-18}$ s) laser pulses- *attoheat thermal wave phenomena.* The attosecond is shorter than all known relaxation times for thermal processes in matter. In that case the duration of the injected pulse is so short that the basic assumption for the applications of the parabolic equations is violated.In the paper we formulate the *theoretical framework* for the ultrashort laser pulse interaction with matter. The quantum thermal Cattaneo and Proca equations are the pillars of the new description of the attheat processes

2. The model equation

Dynamical processes are commonly investigated using laser pump-probe experiments with a pump pulse exciting the system of interest and a second probe pulse tracking is temporal evolution. As the time resolution attainable in such experiments depends on the temporal definition of the laser pulse, pulse compression to the attosecond domain is a recent promising development.

After the standards of time and space were defined the laws of classical physics relating such parameters as distance, time, velocity, temperature are assumed to be independent of accuracy with which these parameters can be measured. It should be noted that this assumption does not enter explicitly into the formulation of classical physics. It implies that together with the assumption of existence of an object and really independently of any measurements (in classical physics) it was tacitly assumed that *there was a possibility of an unlimited increase in accuracy of measurements.* Bearing in mind the "atomicity" of time i.e. considering the smallest time period, the Planck time, the above statement is obviously not true. Attosecond laser pulses we are at the limit of laser time resolution.

With attosecond laser pulses belong to a new Nano – World where size becomes comparable to atomic dimensions, where transport phenomena follow different laws from that in the macro world. This first stage of miniaturization, from $10^{-3}$ m to $10^{-6}$ m is over and the new one, from $10^{-6}$ m to $10^{-9}$ m just beginning. The Nano – World is a quantum world with all the predicable and non-predicable (yet) features.

In this paragraph, we develop and solve the quantum relativistic heat transport equation for Nano – World transport phenomena where external forces exist.

In monograph [7] the new theretical framework for the study of the ultra-short laser pulses interaction with matter was developed, and the hyperbolic equation for temperature field *T(x,t)* was obtained:

$$\frac{1}{\left(\frac{1}{3}v_F^2\right)}\frac{\partial^2 T}{\partial t^2} + \frac{1}{\tau\left(\frac{1}{3}v_F^2\right)}\frac{\partial T}{\partial t} = \nabla^2 T \, , \qquad (2.1)$$

where *T* denotes the temperature, *τ* the relaxation time for the thermal disturbance of the fermionic system, and $v_F$ is the Fermi velocity.

In what follows we develop the new formulation of the HHT, considering the details of the two fermionic systems: electron gas in metals and the nucleon gas.

For the electron gas in metals, the Fermi energy has the form [1]

$$E_F^e = (3\pi)^2 \frac{n^{2/3}\hbar^2}{2m_e}, \qquad (2.2)$$

where $n$ denotes the density and $m_e$ electron mass. Considering that

$$n^{-1/3} \sim a_B \sim \frac{\hbar^2}{me^2}, \qquad (2.3)$$

and $a_B$ = Bohr radius, one obtains

$$E_F^e \sim \frac{n^{2/3}\hbar^2}{2m_e} \sim \frac{\hbar^2}{ma^2} \sim \alpha^2 m_e c^2, \qquad (2.4)$$

where $c$ = light velocity and $a$ = 1/137 is the fine-structure constant for electromagnetic interaction. For the Fermi momentum $p_F$ we have

$$p_F^e \sim \frac{\hbar}{a_B} \sim \alpha m_e c, \qquad (2.5)$$

and, for Fermi velocity $v_F$,

$$v_F^e \sim \frac{p_F}{m_e} \sim \alpha c. \qquad (2.6)$$

. Considering formula (2.6), Eq. (2.1) can be written as

$$\frac{1}{c^2}\frac{\partial^2 T}{\partial t^2} + \frac{1}{c^2\tau}\frac{\partial T}{\partial t} = \frac{\alpha^2}{3}\nabla^2 T. \qquad (2.7)$$

As seen from (2.7), the HHT equation is a relativistic equation, since it takes into account the finite velocity of light.

For the nucleon gas, Fermi energy equals

$$E_F^N = \frac{(9\pi)^{2/3}\hbar^2}{8mr_0^2}, \qquad (2.8)$$

where $m$ denotes the nucleon mass and $r_0$, which describes the range of strong interaction, is given by

$$r_0 = \frac{\hbar}{m_\pi c}, \qquad (2.9)$$

wherein $m_\pi$ is the pion mass. From formula (2.9), one obtains for the nucleon Fermi energy

$$E_F^N \sim \left(\frac{m_\pi}{m}\right)^2 mc^2. \qquad (2.10)$$

In analogy to the Eq. (2.4), formula (2.10) can be written as

$$E_F^N \sim \alpha_s^2 mc^2, \qquad (2.11)$$

where $\alpha_s = \frac{m_\pi}{m} \cong 0.15$ is the fine-structure constant for strong interactions. Analogously, we obtain the nucleon Fermi momentum

$$p_F^e \sim \frac{\hbar}{r_0} \sim \alpha_s mc \qquad (2.12)$$

and the nucleon Fermi velocity

$$v_F^N \sim \frac{pF}{m} \sim \alpha_s c, \qquad (2.13)$$

and HHT for nucleon gas can be written as

$$\frac{1}{c^2}\frac{\partial^2 T}{\partial t^2} + \frac{1}{c^2\tau}\frac{\partial T}{\partial t} = \frac{\alpha_s^2}{3}\nabla^2 T. \qquad (2.14)$$

In the following, the procedure for the discretization of temperature $T(\vec{r},t)$ in hot fermion gas will be developed. First of all, we introduce the reduced de Broglie wavelength

$$\lambda_B^e = \frac{\hbar}{m_e v_h^e}, \qquad v_h^e = \frac{1}{\sqrt{3}} \alpha c,$$
$$\lambda_B^N = \frac{\hbar}{m v_h^N}, \qquad v_h^N = \frac{1}{\sqrt{3}} \alpha_s c, \qquad (2.15)$$

and the mean free paths $\lambda_e$ and $\lambda_N$

$$\lambda^e = v_h^e \tau^e, \qquad \lambda^N = v_h^N \tau^N. \qquad (2.16)$$

In view of formulas (2.15) and (2.16), we obtain the HHC for electron and nucleon gases

$$\frac{\lambda_B^e}{v_h^e} \frac{\partial^2 T}{\partial t^2} + \frac{\lambda_B^e}{\lambda^e} \frac{\partial T}{\partial t} = \frac{\hbar}{m_e} \nabla^2 T^e, \qquad (2.17)$$

$$\frac{\lambda_B^N}{v_h^N} \frac{\partial^2 T}{\partial t^2} + \frac{\lambda_B^N}{\lambda^N} \frac{\partial T}{\partial t} = \frac{\hbar}{m} \nabla^2 T^N. \qquad (2.18)$$

Equations (2.17) and (2.18) are the hyperbolic partial differential equations –damped wave equation which are the master equations for heat propagation in Fermi electron and nucleon gases. In the following, we will study the quantum limit of heat transport in the fermionic systems. We define the quantum heat transport limit as follows:

$$\lambda^e = \lambda_B^e, \qquad \lambda^N = \lambda_B^N. \qquad (2.19)$$

In that case, Eqs. (2.17) and (2.18) have the form

$$\tau^e \frac{\partial^2 T^e}{\partial t^2} + \frac{\partial T^e}{\partial t} = \frac{\hbar}{m_e} \nabla^2 T^e, \qquad (2.20)$$

$$\tau^N \frac{\partial^2 T^N}{\partial t^2} + \frac{\partial T^N}{\partial t} = \frac{\hbar}{m} \nabla^2 T^N, \qquad (2.21)$$

where

$$\tau^e = \frac{\hbar}{m_e (v_h^e)^2}, \qquad \tau^N = \frac{\hbar}{m (v_h^N)^2}. \qquad (2.22)$$

Equations (2.20) and (2.21) define the master equation for quantum wave heat transport (QHT). Having the relaxation times $\tau^e$ and $\tau^N$, one can define the "pulsations" $\omega_h^e$ and $\omega_h^N$

$$\omega_h^e = (\tau^e)^{-1}, \qquad \omega_h^N = (\tau^N)^{-1}, \qquad (2.23)$$

or

$$\omega_h^e = \frac{m_e (v_h^e)^2}{\hbar}, \qquad \omega_h^N = \frac{m (v_h^N)^2}{\hbar},$$

i.e.,

$$\omega_h^e \hbar = m_e (v_h^e)^2 = \frac{m_e \alpha^2}{3} c^2,$$
$$\omega_h^N \hbar = m (v_h^N)^2 = \frac{m \alpha_s^2}{3} c^2. \qquad (2.24)$$

The formulas (2.24) define the Planck-Einstein relation for heat quanta $E_h^e$ and $E_h^N$

$$E_h^e = \omega_h^e \hbar = m_e (v_h^e)^2,$$
$$E_h^N = \omega_h^N \hbar = m_N (v_h^N)^2. \qquad (2.25)$$

The heat quantum with energy $E_h = \hbar \omega$ can be named the *heaton*, in complete analogy to the *phonon, magnon, roton*, etc. For $\tau^e, \tau^N \to 0$, Eqs. (2.20) and (2.24) are the Fourier equations with quantum diffusion coefficients $D^e$ and $D^N$

$$\frac{\partial T^e}{\partial t} = D^e \nabla^2 T^e, \qquad D^e = \frac{\hbar}{m_e}, \qquad (2.26)$$

$$\frac{\partial T^N}{\partial t} = D^N \nabla^2 T^N, \qquad D^N = \frac{\hbar}{m}. \qquad (2.27)$$

The quantum diffusion coefficients $D^e$ and $D^N$ were introduced for the first time by E. Nelson.

For finite $\tau^e$ and $\tau^N$, for $\Delta t < \tau^e$, $\Delta t < \tau^N$, Eqs. (2.20) and (2.21) can be written as

$$\frac{1}{(v_h^e)^2}\frac{\partial^2 T^e}{\partial t^2} = \nabla^2 T^e, \qquad (2.28)$$

$$\frac{1}{(v_h^N)^2}\frac{\partial^2 T^N}{\partial t^2} = \nabla^2 T^N. \qquad (2.29)$$

Equations (2.28) and (2.29) are the wave equations for quantum heat transport (QHT). For Δt > τ, one obtains the Fourier equations (2.26) and (2.27).

In what follows, the dimensionless form of the QHT will be used. Introducing the reduced time $t'$ and reduced length $x'$,

$$t' = t/\tau, \qquad x' = \frac{x}{v_h \tau}, \qquad (2.30)$$

one obtains, for QHT,

$$\frac{\partial^2 T^e}{\partial t^2} + \frac{\partial T^e}{\partial t} = \nabla^2 T^e, \qquad (2.31)$$

$$\frac{\partial^2 T^N}{\partial t^2} + \frac{\partial T^N}{\partial t} = \nabla^2 T^N. \qquad (2.32)$$

and, for QFT,

$$\frac{\partial T^e}{\partial t} = \nabla^2 T^e, \qquad (2.33)$$

$$\frac{\partial T^N}{\partial t} = \nabla^2 T^N. \qquad (2.34)$$

Conclusions

In this paper the quantum Cattaneo equation was proposed. It was shown that for laser pulse with time duration $\Delta t \ll \tau$ the quantum Cattaneo equation is the quantum wave equation. The wave velocity is equal v=αc, where α is the fine structure constant and c is the light velocity